\documentclass[4p,authoryear,12pt]{elsarticle}

\usepackage{amssymb}
\usepackage{amsmath}
\usepackage{siunitx}
 \usepackage{lineno}
 
 \usepackage{multirow}

\journal{Ultrasound in Medicine and Biology}

\begin{document}

\begin{frontmatter}

%% Title


\title{Acoustic measurements of nucleus size distribution at the cavitation threshold}

%% Authors and addresses/affiliations

%% use the corref command within \author for corresponding author footnotes; note the corresponding author can be any of the authors, but one author must be designated; here the third author has been arbitrarily designated as the corresponding author as an example.



\author[Affil1]{Lauren Mancia \corref{cor1}}
\author[Affil2]{Mauro Rodriguez}
\author[Affil3]{Jonathan Sukovich}
\author[Affil3]{Scott Haskell}
\author[Affil3]{Zhen Xu}
\author[Affil1]{Eric Johnsen}
\address[Affil1]{Department of Mechanical Engineering, University of Michigan, Ann Arbor, Michigan, USA}
\address[Affil2]{Division of Engineering and Applied Science, California Institute of Technolog, Pasadena, California, USA}
\address[Affil3]{Department of Biomedical Engineering, University of Michigan, Ann Arbor, Michigan, USA}
% Replace capitalized text with the appropriate information (use standard capitalization rules for your text, not all capitals.
\cortext[cor1]{Corresponding Author: Lauren Mancia, lamancha@umich.edu}

\begin{abstract}
Understanding the acoustic cavitation threshold is essential for minimizing cavitation bioeffects in diagnostic ultrasound and for controlling cavitation--mediated tissue ablation in focused ultrasound procedures.  The homogeneous cavitation threshold is an intrinsic material property of recognized importance to a variety of applications requiring cavitation control. However, acoustic measurements of the cavitation threshold in water differ from those predicted by classical nucleation theories. This persistent discrepancy is explained by combining novel methods for acoustically nucleating single bubbles at threshold with numerical modeling to obtain a nucleus size distribution consistent with first--principles estimates for ion--stabilized nucleii. We identify acoustic cavitation at threshold as a reproducible subtype of heterogeneous cavitation with a characteristic nucleus size distribution. Knowledge of the nucleus size distribution could inspire new approaches for achieving cavitation control in water, tissue, and a variety of other media.
\end{abstract}


\end{frontmatter}

%% Do not remove the page break here.
\pagebreak

%\linenumbers

%% MAIN TEXT INSTRUCTIONS

%% For all sections, subsections, and subsubsections, use the '*' to remove numbering, as demonstrated below.

%% Commands for figures and tables should not be included in the main body of the submitted version of this file (e.g. the figure and tabular environments).  Figure captions should be listed in this file, as shown below.  Tables and Table captions should be listed as a separate section at the end of this file, as shown below.  Many authors may wish to include figures and tables within the main text of their document will preparing their manuscript.  This may be done, however please comment out any of the lines prior to submission.

%% Because the Elsevier editorial process does not allow for the figure and tabular environments in the submitted document, you will be unable to use autonumbering (i.e. \label and \ref) for figures and tables. 

%%  If long equations are used in the document, authors should use a two column format to make sure that the equations will break at approximately the right places.  To do this, replace the class option 'review' with the following two class options '3p' and 'twocolumn'.  Keep in mind that the column width produced in '3p' is slightly narrower than the final printer version.  After inserting the appropriate line breaks in your equation, change the '3p' option back to 'review'.

%% For citations, use the commands \citep and \citet

%% BEGIN MAIN TEXT

%%%%%%%%%%% INTRODUCTION
\section*{Introduction}
\label{intro}
Understanding the acoustic cavitation threshold is essential for the mitigation and control of cavitation bioeffects in diagnostic ultrasound \citep{church2002spontaneous} and for cavitation control in therapeutic ultrasound procedures \citep{bader2019whom}. Homogeneous cavitation occurs when a medium spontaneously ruptures under a tensile (negative) pressure exceeding its 
tensile strength \citep{leighton2012acoustic}. Acoustic measurements of the homogeneous cavitation threshold in 
water range from $-21$ to $-30$ MPa at room temperature 
\citep{herbert2006cavitation,davitt2010water,greenspan1982radiation,bader2019whom}, which are of significantly smaller magnitude than values predicted by classical nucleation theory \citep{debenedetti1996metastable} and measured using microfluidic techniques \citep{ando2012homogeneous}.  Self--ionization of water is a proposed source of ion impurities that destabilize water to cavitation \citep{davitt2010water}. Alternatively, these ions could stabilize preexisting nanoscale gas bubbles against dissolution \citep{akulichev1966hydration} producing \emph{bubbstons} \citep{bunkin1992bubbstons,sankin2003two}. Ion stabilization likely explains the observed longevity of bulk nanobubbles \citep{nirmalkar2018existence,nirmalkar2018interpreting,fang2018formation,zhu2016cleaning,uchida2016effect} and  suggests that acoustic methods could be measuring the onset of heterogeneous cavitation in a subpopulation of ion--stabilized, nanoscale nuclei rather than a genuine homogeneous threshold \citep{maxwell2013probability,sankin2003two}. Nevertheless, the reproducibility of acoustic threshold measurements in water of variable purity implies that this subpopulation of nuclei is highly consistent \citep{maxwell2013probability,borkent2007reproducible,ando2012homogeneous}, \emph{ubiquitous} in water \citep{azouzi2013coherent,davitt2010water}, and \emph{intrinsic} to water and water--based tissues \citep{bader2019whom,bunkin1992bubbstons}. Despite such robust experimental evidence of their existence, these nuclei remain poorly characterized.

Attempts to use fundamental thermodynamics \citep{bunkin1992bubbstons} or nucleation theories to predict a critical or lower--bound cavitation nucleus size at a given temperature \citep{davitt2010water,azouzi2013coherent} provide limited information about the distribution of nuclei in more practical settings, and failure to account for nucleus size variation within a cloud of acoustically--generated bubbles risks neglecting important physics \citep{wang1999effects}. Prior work in heterogeneous cavitation suggests that nucleus sizes follow a lognormal \citep{ben1975bubble,ando2011numerical} or Weibull \citep{wienken2006method} distribution, but it is not clear that these distributions are applicable to nanoscale nuclei present at threshold. While it is possible to measure the size distributions and other characteristics of nanobubbles \citep{nirmalkar2018existence,nirmalkar2018interpreting,fang2018formation,zhu2016cleaning,uchida2016effect,bunkin2014study,bunkin2016formation,jin2007effects}, such studies involve methods that nucleate multiple bubbles simultaneously in water that often contains added ions \citep{zhu2016cleaning,uchida2016effect,bunkin2014study,bunkin2016formation,jin2007effects}.  Though they are likely stabilized by similar physics \citep{akulichev1966hydration}, these nanobubbles are not necessarily representative of the hypothesized nanoscale nuclei present at the acoustic cavitation threshold in deionized water. Moreover, nanoparticle tracking analysis techniques considered most accurate for measuring nanobubble size distributions \citep{nirmalkar2018existence} have detection limits in the tens of nanometers \citep{filipe2010critical}--larger than estimated sizes for acoustic threshold nuclei \citep{maxwell2013probability,bader2019whom} or for preexisting bubbstons in very dilute solutions \citep{bunkin1992bubbstons}. Finally, previous acoustic methods used to infer threshold nucleus sizes for water and other liquids have also been limited by an inability to track individual bubbles from their points of inception \citep{maxwell2013probability,sankin2003two}. An alternative method adapts homogeneous nucleation theory to the study of acoustic cavitation in water and tissue \citep{church2002spontaneous,church1993alternative}. This work assumed spontaneous generation of gas bubbles under energetically--favorable conditions and was used to estimate critical nucleus sizes for given sonication conditions. However, all of these methods are limited to inferring a mean or critical nucleus size that gives rise to a single cavitation event at a measured threshold pressure. To date, no study has both distinguished acoustic cavitation at threshold as a highly reproducible subtype of heterogeneous cavitation and provided measured cavitation statistics for the distribution of preexisting nuclei this implies.

Macroscopic cavitation activity in a variety of disciplines is likely affected by such a nucleus size distribution. The nuclei population is known to determine the onset of ultrasound--induced cavitation in water \citep{brotchie2009effect,bader2019whom} and tissue
\citep{maxwell2013probability,vlaisavljevich2016effects,
vlaisavljevich2014histotripsy,vlaisavljevich2015effects} in biomedical ultrasound. The characteristics of intrinsic nuclei are of particular interest to histotripsy, a non-thermal focused ultrasound procedure that uses controlled cavitation to homogenize soft tissue into acellular debris \citep{xu2005controlled,parsons2006pulsed} for a variety of proposed clinical applications \citep{khokhlova2015histotripsy}. Mechanical tissue fractionation in histotripsy requires the creation of a dense cloud of cavitation bubbles at the treatment focus \citep{parsons2007spatial,xu2005controlled}. Given the stochastic nature of cavitation, understanding the conditions required for bubble cloud generation and maintenance are important for treatment monitoring and planning \citep{bader2019whom}. Furthermore, an understanding of the nuclei population in the relatively controlled setting of histotripsy treatments provides useful insight into nucleation in other settings. For example, cavitation inception in blast traumatic brain injuries \citep{salzar2017experimental} and hydrodynamic applications \citep{chatterjee1997towards} is also thought to involve preexisting nuclei. Moreover, assumptions about the characteristics of initial cavities or defects in adhesives \citep{chikina2000cavitation}, metals \citep{wilkerson2016unraveling}, and amorphous solids \citep{singh2016cavitation,guan2013cavitation} are needed to predict cavitation failure of these materials. Given the 
stochastic nature of cavitation phenomena, the ability to characterize and potentially control the
nuclei population in a given medium would be useful to all of these applications \citep{brotchie2009effect,maxwell2013probability,chatterjee1997towards}. 

This study presents measurements of nanoscale cavitation nuclei in water, specifically, for the first time, a complete size distribution of nuclei induced to grow at the acoustic cavitation threshold. Our measurements are made by combining a unique ultrasound system capable of producing a single cavitation bubble at threshold \citep{wilson2019comparative} with validated numerical modeling \citep{estrada2018high}.

%%%%%%%%%%% MATERIALS AND METHODS
\section*{Methods}
\label{MaM}
      
\subsection*{Single--Bubble Experiments}
Single--bubble experiments were previously performed in a study comparing laser-- to ultrasound--generated cavitation in water and gels  
\citep{wilson2019comparative}, and we leverage the data sets from the water experiments. In brief, water is deionized to a resistivity of 18 megaohms, filtered to $2$ $\mu$m, and 
degassed to $4$ kPa. Experiments use a spherical acoustic array containing $16$ focused transducer elements with a central frequency of $1$ MHz that is capable of generating a single cavitation bubble with a well--characterized pressure waveform. Such control ensures that energy input to grow the bubble is known for a given nucleus size. Single bubbles are nucleated with a probability of $0.5$ using a 1.5--cycle 
acoustic pulse which has a single rarefactional pressure half--cycle with an amplitude of $-24$ MPa. This value is consistent with measurements obtained by our group and others using variable acoustic waveforms and water purity \citep{herbert2006cavitation,maxwell2013probability,vlaisavljevich2016effects}. Images of the bubbles through a single cycle of growth and collapse are obtained using a high--speed camera with 
a multi--flash--per--camera--exposure technique \citep{sukovich2020cost}. This technique generates images of nested, concentric bubbles which are 
differentiated using brightness thresholding and edge detection. Bubble radii are measured at individual flash points by applying a least squares circle fit to their detected boundaries. For all experiments, the magnitudes of the spatial and temporal resolution uncertainties are $4.3$ $\mu$m and $\leq$1.25 $\mu$s, respectively.

\begin{figure*}
\begin{center}
\begin{minipage}{\textwidth}
\includegraphics[width=\textwidth]{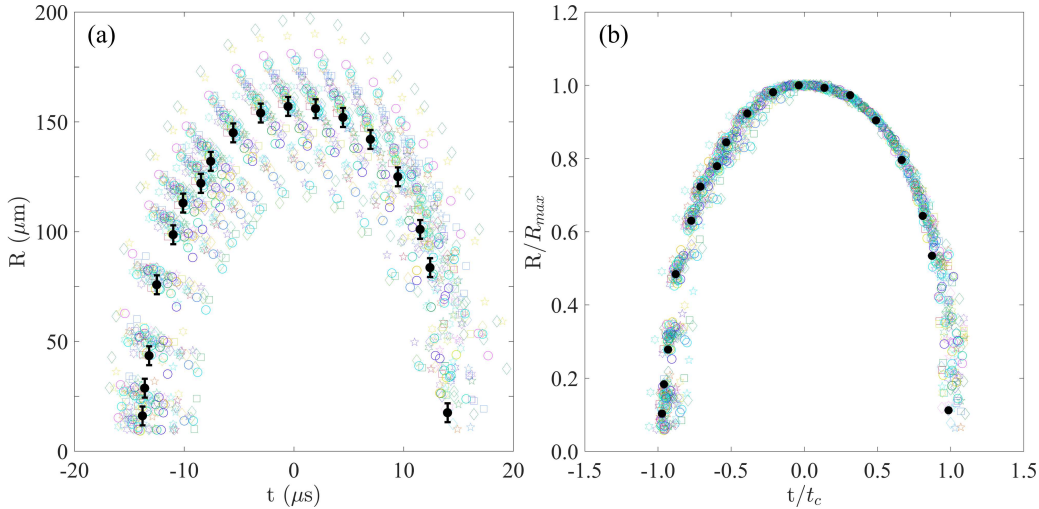}
  \caption{(a) Radius vs.\ time measurements from 88 experiments in water degassed to $4$ kPa. Clustering of data sets is a consequence of aligning all data such that the maximum radii occur at $t = 0$. A single data set is shown in black with spatial resolution error bars. (b) Data sets scaled by maximum radius and collapse time.}
 \label{fig:NucIntro}  
\end{minipage}
 \end{center}
\end{figure*}

Aggregate radius vs.\ time data sets for 88 acoustically--nucleated single--bubble experiments in water degassed to $4$ are shown in Figure 1(a). Although all of the bubbles represented in the curves in Fig.\ 1(a) were generated under equivalent experimental conditions, significant differences between individual experiments are evident.  Black points are a single representative data set, and error bars correspond to uncertainty associated with limitations in spatial resolution. Given that most of the other data sets fall outside of these error bars, it can be concluded that error associated with limitations in spatial resolution does not explain the data spread. Figure 1(b) shows that the data sets collapse when scaled by the measured maximum radius and collapse time, $t_c = 0.92R_{max}\sqrt{\rho_{\infty}/P_{\infty}}$, where $R_{max}$ is the maximum bubble radius of a given data set, $\rho_{\infty}$ is liquid density, and $P_{\infty}$ is the far--field liquid pressure (both constants given in Table 1). Thus, all experiments are governed by the same physics with exceedingly small uncertainty once appropriate scaling addresses uncertainties in initial conditions. In particular, the energy delivered by the ultrasound pulse to the nucleus grows the bubble as the bubble expands to maximum radius against its surroundings. That initial energy is proportional to the nucleus volume and ultrasound pressure amplitude. Given the negligible error in pressure amplitude measurement and its consistency with previous studies \citep{davitt2010water}, we submit that the data spread is due to different nanoscale nucleus sizes corresponding to each experiment. Validated numerical methods can then be used in combination with experimental data to infer these unknown nucleus sizes.

\subsection*{Theoretical Model}      
We simulate the dynamics of a single spherical, homobaric air bubble in water.  To account for near--field compressibility effects, radial bubble dynamics are described by the Keller--Miksis equation \citep{keller1980bubble}:
\small
\begin{align}
\label{KM}
\begin{split}
&\left(1 -\frac{\dot{R}}{c_{\infty}}\right)R\ddot{R}+\frac{3}{2}
\left(1 - \frac{\dot{R}}{3c_{\infty}}\right)\dot{R}^2 = \\
&\frac{1}{\rho_{\infty}}\left(1 +\frac{\dot{R}}{c_{\infty}}
+\frac{R}{c_{\infty}}\frac{d}{dt}\right)\Biggl[p_B-\\
&\Biggl(p_{\infty}+ p_{f}\Biggl(t + \frac{R}{c_{\infty}}\Biggr)\Biggr)-\frac{2\sigma}{R}+J\Biggr],
\end{split}
\end{align}
\normalsize
where $R$ is the bubble radius, $c_{\infty}$ and
$\rho_{\infty}$ are the constant sound speed and density of the
medium, respectively. The surface tension, $\sigma$ and viscosity, $\mu$ are constants for water at $25\,^{\circ}\mathrm{C}$. These parameters and others that remain constant for all simulations are given in Table 1. 

Heat transfer effects are considered by solving for temperature fields inside and outside of the bubble.  The time derivative of the internal bubble pressure, $p_B(t)$ couples the Keller--Miksis equation (Eq.\ \ref{KM}) to the energy equation for air inside the bubble:

\begin{align} \label{eq:pB}
\dot{p}_B=\frac{3}{R}\left((\kappa -1)K\frac{\partial T}{\partial r}\bigg|_{R}-\kappa p_B\dot{R}\right)
\end{align}

\begin{align} \label{eq:kappa}
\begin{split}
\frac{\kappa}{\kappa-1}\frac{p_B}{T}\left[\frac{\partial T}{\partial t}+\frac{1}{\kappa p_B}\left((\kappa -1)K\frac{\partial T}{\partial r}-\frac{r\dot{p}_B}{3}\right)\frac{\partial T}{\partial r}\right]\\
=\dot{p}_B + \frac{1}{r^2}\frac{\partial}{\partial r}\left(r^2K\frac{\partial T}{\partial r}\right),
\end{split}
\end{align}

\noindent
where $T(r,t)$ is the temperature field of air inside the bubble, which has a ratio of specific heats $\kappa$. The air has a thermal conductivity given by $K = K_AT + K_B$, where constants $K_A$ and $K_B$ were determined empirically for air \citep{prosperetti1988nonlinear}. The initial pressure inside the bubble is $p_B(0) = p_{\infty} + 2S/R$. A boundary condition is prescribed for the center of the bubble: $\nabla T=0$ at $r=0$. The bubble wall boundary condition is simplified to $T(R) = T_{\infty}$ under the assumption that the water remains at its constant ambient temperature through the single cycle of bubble growth and collapse considered in each experiment \citep{prosperetti1991thermal,estrada2018high}. 

The far--field pressure is the sum of the ambient pressure, 
$p_{\infty}$ and time--varying incident pulse, $p_f(t)$:
\begin{align}
    p_f(t)= 
\begin{cases}
    p_A\left(\frac{1+\cos[\omega(t-\delta)]}{2}\right)^n,&
      |t-\delta|\leq \frac{\pi}{\omega},\\ 0, & |t-\delta|>
      \frac{\pi}{\omega}.
\end{cases}
\end{align} 
The pressure amplitude, $p_A=-24$ MPa and frequency, $f=1$
MHz ($\omega = 2\pi f$) are approximated from experimental waveform measurements while the time delay, $\delta = 5$ $\mu$s and fitting parameter, $n$ are chosen as in previous studies \citep{vlaisavljevich2014histotripsy,mancia2017predicting}. Based on the notion of a cavitation threshold, cavitation only occurs if a sufficiently large tension is applied. The experimental waveform consists of several cycles, and only the peak negative portion is beyond threshold. An analytic approximation of this peak negative pressure portion of a raw experimental waveform \citep{mancia2019modeling} is valid in this near ideal case of inertial cavitation in which resonant frequency of the bubble is much less than the driving frequency of the waveform.

\begin{table}[t]
\centering
%\captionsetup{justification=centering}
\caption{Constant Parameters}
\label{table:Constants}
\begin{tabular}{c c} 
 \hline\hline
 Parameter & Value \\
\hline 
  $p_A$ & \SI{-24} {\MPa}\\
  $f$ & \SI{1} {\MHz}\\
  $n$ & 3.7 \\
  $\delta$ & \SI{5} {\mu \s}\\
  $S$ & \SI{72} {\mN \per {\meter}}\\
  $c_{\infty}$ & \SI{1496} {\m \per {\sec}}\\
  $\rho_{\infty}$ & \SI{1000} {\kg \per {\meter \cubed}}\\
  $_{\infty}$ & \SI{101.325} {\kPa}\\
  $T_{\infty}$ & \SI{25} {\celsius} \\
  $\kappa$ & 1.4\\ 
  $K_A$ & \SI{5.28e-5}{\W \per {\meter \K \squared}}\\
  $K_B$ &  \SI{1.165e-2}{\W \per {\meter \K}}\\
  \hline\hline 
\end{tabular}
\end{table}

\subsection*{Problem Setup}
The equations are nondimensionalized
\citep{warnez2015numerical} using the initial bubble radius, $R_0$, water density, $\rho_{\infty}$, equilibrium pressure of the bubble contents, $p_{0} = p_{\infty} + 2\sigma/R$, and 
far--field temperature, $T_{\infty}$ to define a characteristic speed, $u_c = \sqrt{p_0/{\rho_{\infty}}}$ and dimensionless 
parameters: Reynolds number, Re $= \rho_{\infty} u_cR_0/\mu$, Weber number, We $=p_0R_0/2S
$, dimensionless sound speed, $C=c_{\infty}/u_c$, and $\chi = 
T_{\infty}K_M/p_0R_0u_c$. A variable--step, variable--order solver based on numerical differentiation formulas (MATLAB \textit{ode15s}) is used for numerical time marching \citep{shampine1997matlab,shampine1999solving}. Equations are 
integrated over a dimensional time span of $t = [0,50]$ in 
microseconds; results are then time--shifted so that the maximum bubble radius occurs at $t = 0$. Using numerical methods described by 
\citep{warnez2015numerical}, the spatial derivatives in the energy equation are discretized on a mesh of $N_s
+1$ points in $r$-space \citep{prosperetti1988nonlinear} inside the 
bubble and computed using a spectral collocation method \citep{warnez2015numerical}. Results are sufficiently converged when simulations use $N_s=30$ 
points inside the bubble. A more detailed treatment of the derivation and numerical implementation of this model can be found in the literature \citep{prosperetti1988nonlinear,kamath1993theoretical,barajas2017effects,
warnez2015numerical}.

%%%%%%%%%%% Results
\section*{Results}
\label{Results}

To construct the nucleus size distribution, we hypothesize that cavitation nuclei exist as stabilized nanobubbles \citep{maxwell2013probability,sankin2003two} and seek to determine the minimum nucleus size, $R_0^*$ required for cavitation growth at a given threshold pressure. Based on past work on the acoustic cavitation threshold \citep{maxwell2013probability,vlaisavljevich2015effects,vlaisavljevich2016effects}, the threshold pressure is fixed at its measured value of $-24$ MPa for all simulations. Figure 2(a) shows the simulation maximum bubble radius as a function of nucleus size under this tensile pressure. Bubble growth is negligible until a minimum nucleus size of $R_0^*=$ $2.32$ nm is reached. Because the time of the tensile pulse is much longer than the timescale of the bubble, the quasistatic assumption holds, and the minimum nucleus size can be calculated from the Blake threshold \citep{leighton2012acoustic}. The minimum pressure amplitude needed to cause explosive growth of a bubble with initial radius $R_0$ is given by:

\begin{align}\label{eq:BlakeP}
P_B = P_{\infty} + \frac{8\sigma}{9}\sqrt{\frac{3\sigma}{2(P_{\infty} +2\sigma/R_0)R_0^3}},
\end{align} 

\noindent
where $P_B$ is the Blake threshold, $P_{\infty} = 101.325$ kPa is the ambient pressure of the surrounding fluid, and $\sigma =  0.072$ N/m is the surface tension of water at $25^{\circ}$C.  Assuming $2\sigma/R_0 \gg P_0$ for these nanoscale nuclei gives rise to a simplified expression for $R_0$:

\begin{align}\label{eq:R0}
R_0 = \frac{4\sigma}{3\sqrt{3}}\bigg(\frac{1}{P_B - P_{\infty}}\bigg).
\end{align} 

\noindent 
In the present case, the Blake threshold pressure is equivalent to the measured threshold pressure: $P_B = 24$ MPa. Substituting the other physical constants into Eq.\ \ref{eq:R0} gives $R_0 = 2.32$ nm $ = R_0^*$, which is the minimum bubble radius that will grow when exposed to the measured Blake threshold pressure \citep{walton1984sonoluminescence}. For comparison, previous studies estimate the minimum radii of stabilized nanoscale nuclei to be approximately $2.0$ nm \citep{bunkin1992bubbstons} from first principles and $2.5$ nm \citep{maxwell2013probability} from bubble dynamics simulations. In contrast, critical nucleus volumes obtained using homogeneous nucleation theories correspond to radii of $1.3$ nm \citep{davitt2010water} and $0.88$ nm \citep{azouzi2013coherent} at $300$ K.      

\begin{figure}[t]
\centering
  \includegraphics[width=0.57\linewidth]{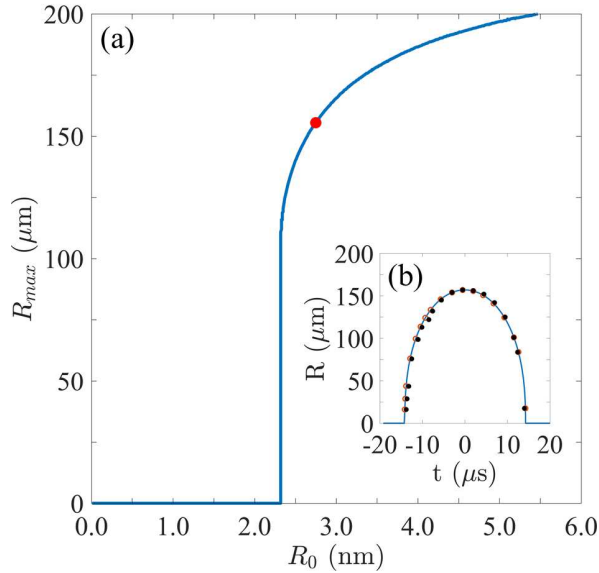}
  \caption{(a) Simulation $R_{max}$ vs.\ nucleus size. (b) A representative data set in black. The simulation (blue trace) initialized with $R_0$ indicated by solid red point in (a) optimizes the normalized rms error between experiment and simulation nearest neighbors (red open points).}
  \label{fig:Thresh0}
\end{figure}

The complete nucleus size distribution is created by varying the 
$R_0$ used to initialize simulations over a range of $2.32 - 6.00$ nm for each experimental data set. A nearest neighbors algorithm with a standardized Euclidean distance metric is then used to identify simulation points closest to experimental data points. The nucleus distribution consists of $R_0$ values that optimize the normalized 
root--mean--squared (rms) error between individual data points of a given experimental realization and their simulation nearest neighbors. The average normalized rms error for these data sets is $0.98$ (with $1.00$ implying a perfect fit). Figure 2(b) shows the representative data set from Figure 1. The simulation initialized with $R_0 = 2.78$ nm (indicated by the red 
point in Figure 2(a)) optimizes the normalized rms error between the experimental data (black points) and the nearest neighbors on the simulation trace (red open points).  In this case, the normalized rms error is $0.98$, which is equivalent to the mean error for all data sets. This procedure is followed for each data set to obtain $R_0$ values optimizing the normalized rms error.     

\begin{figure}[t]
\centering
  \includegraphics[width=0.57\linewidth]{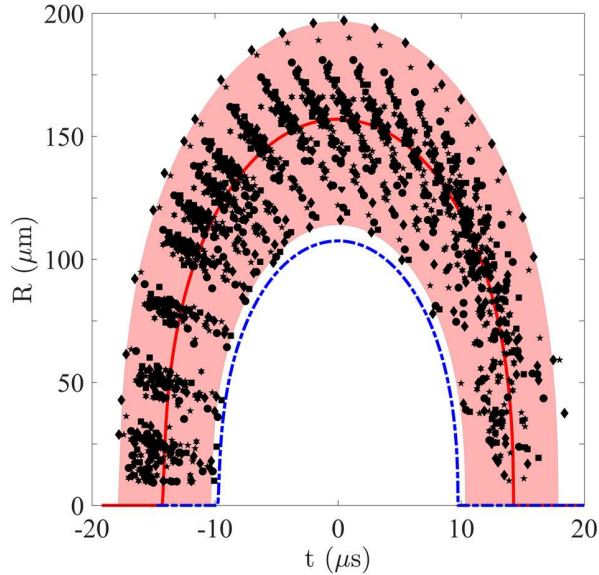}
  \caption{Experimental data from Fig.\ 1. Shaded region is bounded by simulations initialized with the smallest and largest nucleii. The dark red trace is the simulation corresponding to the mean nucleus size. The dashed blue trace is the simulation initialized with $R_0^*$.}
  \label{fig:R0Bands}
\end{figure}

\begin{figure}
\centering
  \includegraphics[width=0.57\linewidth]{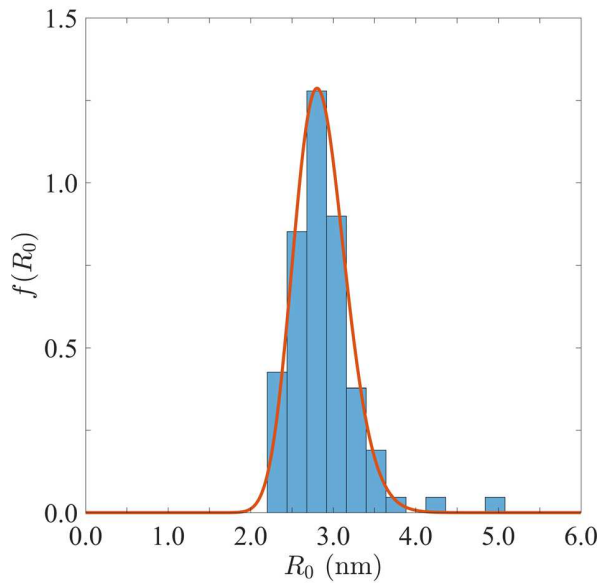}
  \caption{Nucleus size distribution fitted to lognormal pdf (red).}
  \label{fig:R0Dist}
\end{figure}

Simulations initialized with different nucleus sizes effectively bound the experimental data sets as shown in Figure 3.  Aggregate experimental data from Figure 1 are shown in black, and the shaded region is bounded by simulations initialized with the smallest nucleus size, $R_0 = 2.33$ nm and the largest nucleus size, $R_0 = 4.99$ nm that optimize the normalized rms error. The dark red trace is the simulation initialized with the mean nucleus size of $2.88$ nm, and the blue dashed trace is the simulation initialized with the predetermined, lower--bound nucleus, $R_0^* =$ $2.32$ nm. The nucleus size distribution is best approximated by a lognormal probability distribution function (pdf), outlined in red in Figure 4, which has $\sigma = 0.11$ and $\mu = 1.0$. This finding is consistent with previous use of a lognormal distribution to model equilibrium bubble sizes for polydisperse flow based on measured bubble populations in a water tunnel and ocean water \citep{ben1975bubble,ando2011numerical}.

%%%%%%%%%%% DISCUSSION
\section*{Discussion}
\label{Discuss}
The nucleus size distribution is consistent with previous studies which estimated the sizes of ion--stabilized nuclei to be between $2$ and $4$ nm \citep{sankin2003two,bunkin1992bubbstons}, and supports the hypothesis that hydronium ions (e.g. those produced during self--ionization of water) are the ubiquitous impurity responsible for the discrepancy between acoustically--measured and theoretical homogeneous cavitation thresholds \citep{davitt2010water}. The lognormal pdf parallels size distributions measured for larger cavitation bubbles in settings of heterogeneous cavitation \citep{ben1975bubble,ando2011numerical}, and nuclei measured in this study are at least $1$ nm larger than critical nucleus sizes obtained using homogeneous nucleation theories at comparable pressure amplitudes. These findings suggest that acoustic methods, even in highly purified water, are measuring a threshold for heterogeneous rather than homogeneous cavitation. However, consistency in measurements \citep{herbert2006cavitation,davitt2010water,greenspan1982radiation} distinguishes cavitation at the acoustic threshold as a reflection of the nucleii population intrinsic to that medium.

Despite significant differences in waveform and water quality used in previous experiments \citep{herbert2006cavitation,maxwell2013probability}, measured acoustic cavitation thresholds differ from each other and from ours by $< 4$ MPa.  A previous study also notes that their threshold measurements are stable even to the deliberate introduction of impurities \citep{herbert2006cavitation}.   However, there is evidence that gas concentration of the water could give rise to larger nuclei \citep{akulichev1966hydration}. To investigate this possibility, preliminary experiments were performed with water degassed to $40$ kPa ($\sim 40$\% oxygen saturation) instead of to the original $4$ kPa ($\sim 4$\% oxygen saturation) in our original experiments \citep{wilson2019comparative}. The higher gas content had a negligible effect on the measured acoustic cavitation threshold of $-24$ MPa.  From $28$ single--bubble experimental data sets, we inferred a mean nucleus size of $3.60$ nm with nucleus sizes ranging from $2.64$ nm to $5.78$ nm. Although the newly measured nuclei are slightly larger, as expected from the arguments presented in \citep{akulichev1966hydration}, the mean nucleus sizes agree to within $< 1$ nm despite the ten--fold difference in gas content of the water.  These results support previous findings that both the acoustic cavitation threshold is relatively stable to changes in water purity. Future work will further investigate the role of gas content, pH, and additive ions on the acoustic cavitation threshold in water and its associated nucleus size distribution.

Our method for determining the nucleus size distribution infers quantities well below the resolution limits of experiments, but our theory could be strengthened by greater consideration of nanoscale physics. For example, ion interactions could affect the earliest stages of bubble growth when nucleus sizes are nanoscopic. Additionally, nucleation phenomena are highly dependent on surface tension \citep{church2002spontaneous}, and the effective surface tension experienced by a nanoscale nucleus differs from that of the bulk medium \citep{azouzi2013coherent}. Investigation of these effects will be the subject of future work, with molecular dynamics simulations offering the most robust approach. Finally, our work has focused on water given its well--characterized physical properties and an acoustic cavitation threshold that is comparable to that of water--based soft tissues \citep{bader2019whom}. In future work, we intend to extend these results to other liquids and viscoelastic media exhibiting thresholds outside of the typical range for water and water--based tissues \citep{maxwell2013probability}.

%%%%%%%%%%% Conclusions
\section*{Conclusions}
\label{Conclusions}

In summary, this work presents a new approach for using single--bubble experiments and numerical simulations to measure the size distribution of nanoscale cavitation nucleii present at the acoustic cavitation threshold. Recognizing that the leading--order experimental uncertainty lies in the initial nucleii population, the inverse problem for the nucleus size distribution is solved with a single--bubble numerical model. The nucleus size distribution obeys a lognormal pdf ranging from $2.33$ to $4.99$ nm with a mean of $2.88$ nm. Although water is the only medium considered in this study, the methods presented here could be readily extended to predict the intrinsic nucleus size distributions characteristic of other liquids and tissue--like media, thus offering a new avenue for achieving cavitation control in biomedical ultrasound and a variety of other applications.

%%%%%%%%%%% ACKNOWLEDGEMENTS
\section*{Acknowledgements}
\label{Ack}
The authors wish to thank Prof.\ J.\ Brian Fowlkes for helpful discussions. This work was supported by ONR Grant No. N00014-18-1-2625 (under Dr. Timothy Bentley).


%% \section*{}

%%%%%%%%%%% REFERENCES
%% REFERENCE FORMATTING INSTRUCTIONS

%% All bibliography information should be included using a 'thebibliography' environment.  Most authors will find it easiest to create a .bbl file using the commands \bibliographystyle{} and \bibliography{} and then copy and paste the contents of the .bbl file into the .tex file below, but before the figure captions section.  Examples for using the \bibliographystyle and \bibliography commands are listed below.  

%% Do not remove the page break here.
\pagebreak

%% References with bibTeX database, use this to create a .bbl file
\bibliographystyle{UMB-elsarticle-harvbib}
\bibliography{NucBib}

%% References copied and pasted from the .bbl file.  Copy and paste over the following two lines.  When using a bibTeX database to create a .bbl file, comment out the following two lines.
%\begin{thebibliography}{00}
%\end{thebibliography}

%%%%%%%%%%% FIGURE CAPTIONS

%% Include only the figure captions here (not the figures).  Figures are uploaded separately in the online Elsevier Editorial Submission process.

%% Do not remove the page break here.
\pagebreak

\pagebreak

\pagebreak

%\section*{Video Captions}
%
%\begin{description}
%\item[Figure 1:]  TYPE THE CAPTION FOR VIDEO ONE HERE.
%\item[Figure 2:]  TYPE THE CAPTION FOR VIDEO TWO HERE.
%\item[Figure 3:]  TYPE THE CAPTION FOR VIDEO ONE HERE.  CONTINUE THIS LIST FOR ALL OTHER CAPTIONS
%\end{description}

\end{document}